\documentclass[conference]{IEEEtran}
\IEEEoverridecommandlockouts
\usepackage{cite}
\usepackage{amsmath,amssymb,amsfonts}
\usepackage{algorithmic}
\usepackage{graphicx}
\usepackage{textcomp}
\usepackage{xcolor}
\usepackage{multirow}

\def\BibTeX{{\rm B\kern-.05em{\sc i\kern-.025em b}\kern-.08em
    T\kern-.1667em\lower.7ex\hbox{E}\kern-.125emX}}
\begin{document}

\title{Semantic Synchronization for Enhanced Reliability in Communication Systems\\

\thanks{Corresponding author: Chen Dong, e-mail:dongchen@bupt.edu.cn.}\thanks{This work is supported in part by the National Key R$\&$D Program of China under Grant 2022YFB2902102.}
}

\author{\IEEEauthorblockN{Xiaoyi Liu\IEEEauthorrefmark{2}, Haotai Liang\IEEEauthorrefmark{2}, Chen Dong\IEEEauthorrefmark{2}, Xiaodong Xu\IEEEauthorrefmark{2}\IEEEauthorrefmark{3}}
\IEEEauthorblockA{\IEEEauthorrefmark{2} the State Key Laboratory of Networking and Switching Technology, \\
Beijing University of Posts and Telecommunications, Beijing, China}

\IEEEauthorblockA{\IEEEauthorrefmark{3} the Department of Broadband Communication, Peng Cheng Laboratory, Shenzhen, Guangdong, China}
}
\maketitle

\begin{abstract}
As a new communication paradigm, semantic communication has received widespread attention in communication fields. However, since the decoding of semantic signals relies on contextual knowledge, misalignment between the starting position of the semantic signal and the AI-based semantic decoder would prevent source signal recovery and reconstruction. To achieve more precise semantic communication, this study proposes an image-based semantic synchronization method leveraging intrinsic semantic features of image content. Specifically, a shared synchronized image (SyncImg) is encoded into a synchronization vector header at the transmitter and sent to the receiver. The receiver adopts a sliding window semantic decoder combined with classification and template matching methods to locate the synchronization point. Experimental results demonstrate that compared with traditional methods, the proposed method achieves a lower miss detected ratio (MDR) and root-mean-square error (RMSE) under low signal-to-noise ratios, realizing accurate synchronization of semantic signals across different devices.
\end{abstract}


\section{Introduction}
With the rapid development of communications, semantic communication has attracted the attention of researchers as a new communication paradigm. Compared with traditional approaches, semantic communication aims to achieve more precise and intelligent information exchange through conveying semantic content. Research on semantic communication sources has covered text \cite{DeepSC}, speech \cite{DeepSC-S}, image \cite{LSCI}\cite{NTSCC}, video \cite{MDVSC}\cite{20_DVST}, point cloud \cite{PCSC} and extended reality (XR) \cite{XR}. Besides, these systems have demonstrated good performance even in low signal-to-noise ratio (SNR) conditions.
\par However, semantic signals in semantic communication can only be interpreted by AI-based encoders and decoders, where the encoding and decoding processes require considering global contextual information of the source and semantic signals. Misalignment between the semantic signal starting position and decoder prevents source signal reconstruction. Hence, synchronization plays a crucial role in ensuring the regular operation of the semantic decoder. Traditional synchronization methods mainly rely on timestamps or frame numbers, using self-correlation sequences (such as M-sequences) or similar STO synchronization methods in OFDM, where a portion of the data is copied into a cyclic prefix (CP) to make the CP and data correlated, thus achieving synchronization. Currently, there are also some research studies on synchronization methods. For example, a subframe synchronization method based on cyclic prefix self-correlation is proposed in \cite{OFDM_synchronization}, which uses the CP correlation results of multiple symbols to determine the starting point of the subframe for better synchronization. In addition, literature \cite{AI_synchronization} utilizes deep learning methods for signal detection by training a neural network to predict the starting position of the signal. However, applying traditional synchronization methods based on statistical features results in high synchronization costs and difficulties meeting the semantic communication requirement for synchronization localization performance under low transmission rate conditions in low SNR environments. Therefore, designing a synchronization method that can extract and utilize the semantic features of the sequence may improve synchronization accuracy in low SNR environments while maintaining low costs.
\par To achieve more precise semantic communication, especially accurate semantic communication under low SNR conditions, this paper leverages image intrinsic correlation and semantic communication principles to propose an image-based semantic synchronization method for more precise synchronization. Specifically, both the transmitter and receiver share a synchronized image (SyncImg). At the transmitter, a semantic encoder encodes the SyncImg and unfolds it into a one-dimensional vector as the synchronization header, which is then transmitted along with the data. At the receiver, a sliding window approach is employed, with the window size matching the length of the encoding header.  With each sliding unit, the semantic decoder decodes the data within the window, and the reconstructed image is synchronized and positioned using a classification model and template matching.
\par The main contributions of this paper include:
\par 1) To the best of the authors’ knowledge, this paper first discovers and analyzes the impact and reasons for the misalignment of semantic symbols in image-based semantic recovery. Based on this, an image-based semantic synchronization system is designed for accurate synchronization positioning.
\par 2) The proposed semantic communication synchronization method performs well in a low signal-to-noise ratio environment, improving the reliability of low signal-to-noise ratio communication. Experimental results demonstrate that under low signal-to-noise ratio conditions, the proposed method in this paper outperforms traditional methods in metrics such as missed detection rate (MDR) and root mean square error (RMSE).
\par The rest of this paper is organized as follows: Section \uppercase\expandafter{\romannumeral2} introduces the system model. Section \uppercase\expandafter{\romannumeral3} elaborates on the principles of the semantic synchronization method. Section \uppercase\expandafter{\romannumeral4} describes the experimental setup and results analysis, and finally, Section \uppercase\expandafter{\romannumeral5} summarizes the entire paper.

\section{System Model}

\begin{figure}[b]
\centering
\includegraphics[width=\linewidth]{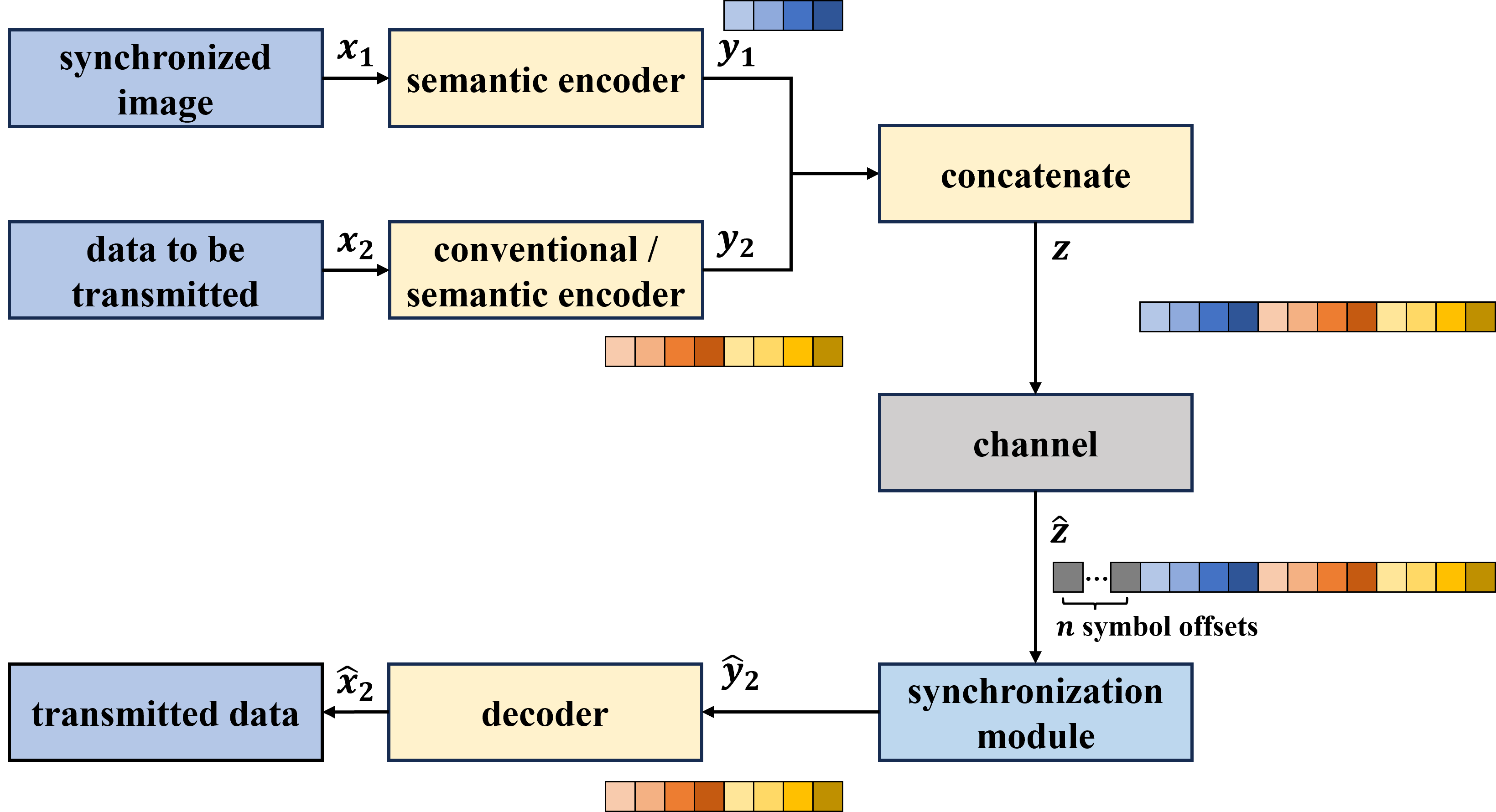}
\caption{The framework of the semantic synchronization system.}
\label{fig1}
\end{figure}

The proposed semantic synchronization framework is illustrated in Fig. \ref{fig1}. First, at the transmitter, the SyncImg \(x_1\) is encoded into \(y_1\) using the semantic encoder:
\begin{equation}
y_1 = SC_{\alpha}(x_1),
\label{eq1}
\end{equation}
where \(SC_\alpha(\cdot)\) represents the semantic encoder with parameter \(\alpha\).

The data to be transmitted, \(x_2\), is encoded into \(y_2\) using an encoder:
\begin{equation}
y_2 = En(x_2),
\label{eq2}
\end{equation}
where \(En(\cdot)\) denotes the encoder function. The encoder can be a different semantic encoder from Eq. (\ref{eq1}) or a conventional one. This study focuses its analysis on the signal $y_1$. The other signal $y_2$ can be either a semantic signal or a signal encoded by a conventional encoder. Then, $y_1$ and $y_2$ are flattened into one-dimensional vectors and concatenated into a sequence $z$. Here, $y_1$ serves as the synchronization header to facilitate the receiver in determining the starting position of data. The flattening rule of $y_1$ will be elaborated in Section \uppercase\expandafter{\romannumeral3}. This paper does not restrict the expansion approach of $y_2$. Subsequently, $z$ is transmitted over the wireless channel. The received signal at the receiver, \(\hat{z}\), can be expressed as:
\begin{equation}
\hat{z} = z + n,
\label{eq3}
\end{equation}
where \(n\) represents additive white Gaussian noise with \(CN(0,\sigma^2)\). 

At the receiver, the synchronization module, which will be introduced in Section \uppercase\expandafter{\romannumeral3}, performs signal detection to locate the starting position of the transmitted data stream accurately. The transmitted data \(\hat{y}_2\) can be expressed as:
\begin{equation}
\hat{y}_2 = Sy(\hat{z}),
\label{eq4}
\end{equation}
where \(Sy(\cdot)\) represents the synchronization method.

Finally, term \(\hat{y}_2\) is fed into the decoder for decoding to obtain \(\hat{x}_2\):
\begin{equation}
\hat{x}_2 = De(\hat{y}_2),
\label{eq5}
\end{equation}
where \(De(\cdot)\) denotes the decoder function.

\section{System Design}

\subsection{Flattening Rule}
\begin{figure}[b]
\centering
\includegraphics[width=\linewidth]{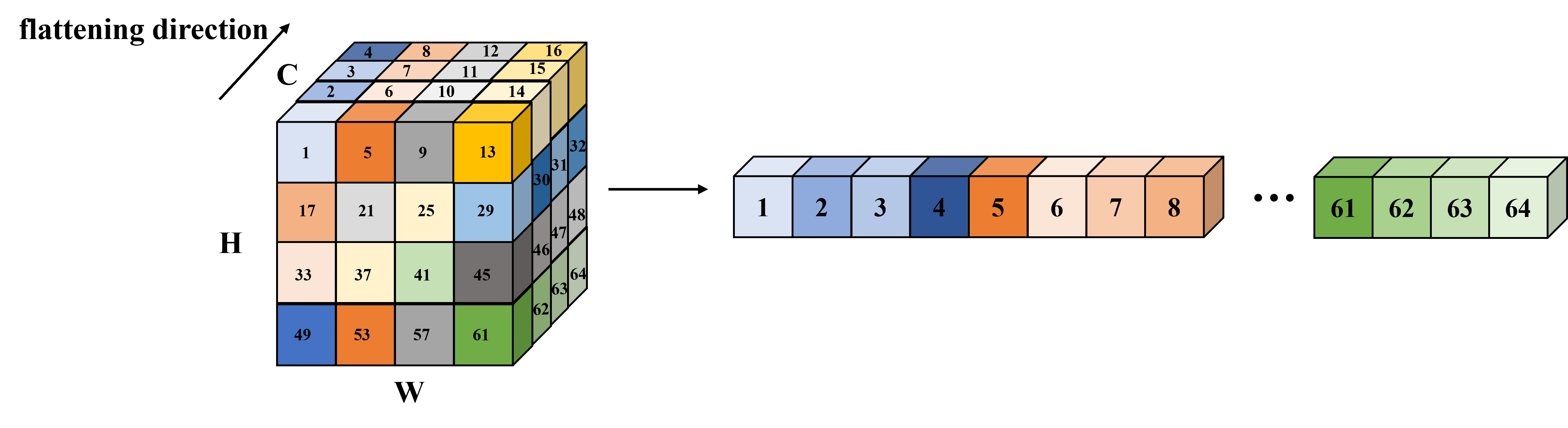}
\caption{The flatten method of the semantic synchronization header.}
\label{fig2}
\end{figure}

In Section \uppercase\expandafter{\romannumeral2}, the encoded semantic vector needs to be flattened into a one-dimensional vector. Assuming that the SyncImg is encoded and represented as $(B, C, H, W)$, where $B$ denotes the batch size, $C$ represents the number of channels, $H$ indicates the tensor height, and $W$ means the tensor width, the flattening process is performed along the channel dimension. For instance, in the specific scenario of $B=1$, $C=4$, $H=4$, and $W=4$, the encoded semantic vector exhibits a structured arrangement, as illustrated on the left side of Fig. \ref{fig2}. The resulting one-dimensional vector, obtained by flattening along the channel direction, is demonstrated on the right side of Fig. \ref{fig2}.

\subsection{Semantic Synchronization Principle}

\begin{figure*}[t]
\centering
\includegraphics[width=140mm]{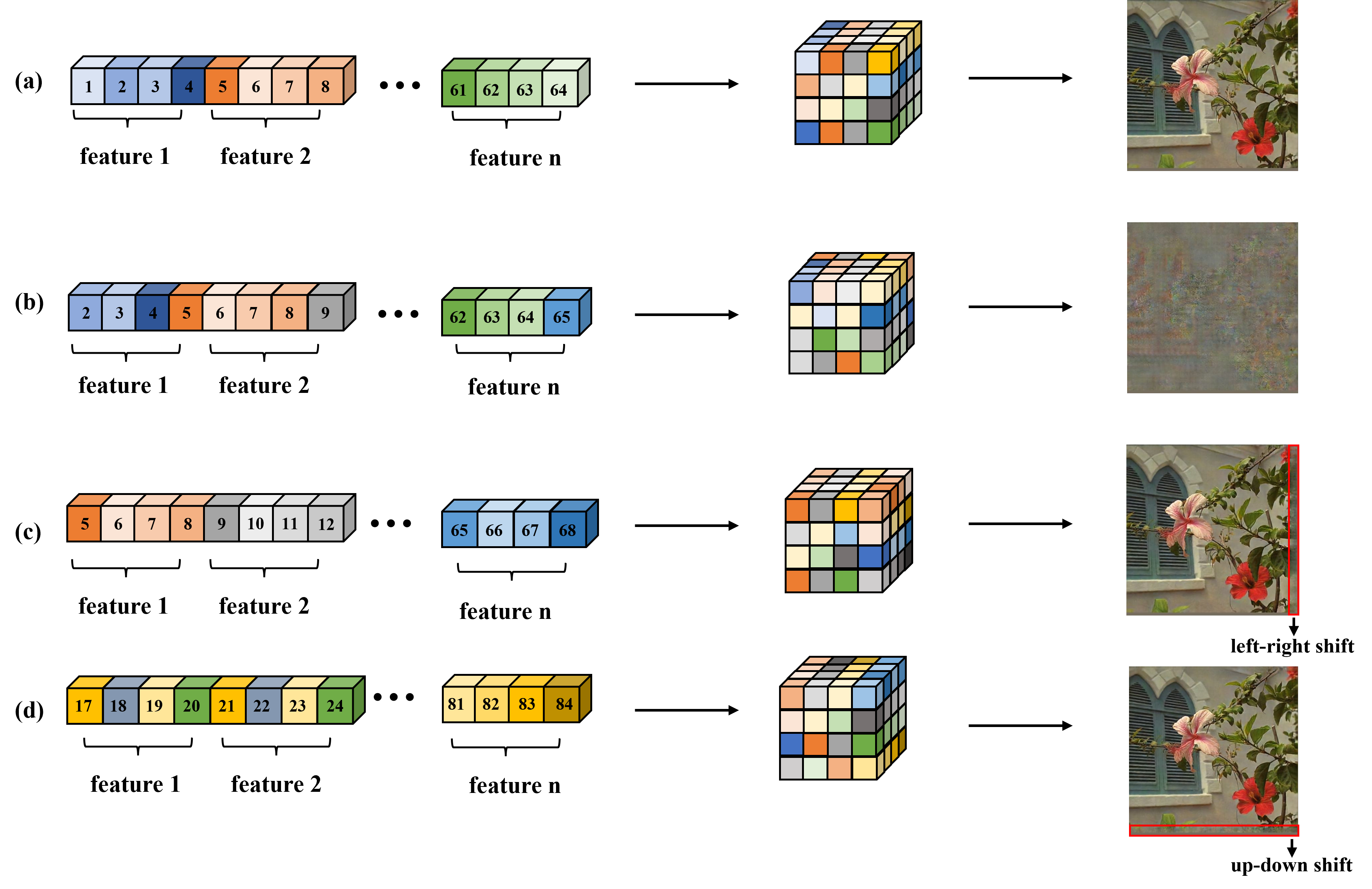}
\caption{The impact of the symbol offset on image reconstruction. A $512\times512$ cropped image from the Kodak dataset \cite{Kodak} is used as an example. (a) Semantic vectors and the image without any symbol offset. (b) Semantic vectors and the image when the semantic symbols experience a symbol offset smaller than $C$, resulting in a noisy reconstructed image. (c) Semantic vectors and the image when the semantic symbols experience a symbol offset of $C \times n$ (where $n$ is a positive integer and $n < W$), causing a left-right shift in the reconstructed image. (d) Semantic vectors and the image when the semantic symbols experience a symbol offset of $C \times W \times m$ (where $m$ is a positive integer), leading to an up-down shift in the reconstructed image.}
\label{fig3}
\end{figure*}

The semantic encoder mentioned in Section \uppercase\expandafter{\romannumeral2} consists of multiple 2D convolutional layers. When performing repeated 2D convolution operations on an image, each convolutional layer generates a new set of semantic feature maps. These semantic feature maps are stacked in the channel direction to form a three-dimensional tensor, where each element represents a feature. Each convolutional kernel contains a set of trainable weight parameters in the channel direction. The convolution kernel can extract feature information in the corresponding order by convolving with the input feature maps. Therefore, the group of vectors in the channel direction can be understood as a set of semantic features for describing certain specific features in the input image.

Based on the above principles, when using the flattening method mentioned earlier, every $C$ symbol represents a set of semantic features. If the sequence undergoes a shift that is not an integer multiple of $C$, the order of the corresponding semantic features of each feature will be disturbed after shifting. Since the semantic decoder and semantic encoder share contextual knowledge, scrambled features will cause decoding errors. Assuming $C=4$ and the sequence shifts one position to the left, as shown in Fig. \ref{fig3}(b), each semantic feature will be disordered, resulting in a noisy image when decoded.

Suppose the sequence undergoes a shift of $C \times n$ (where $n$ is a positive integer less than $W$). In that case,  the relative order between different features changes, but each feature is unaffected. For this reason, the semantic decoder can correctly restore the semantic vector that has undergone the shift of $C \times n$. However, the decoded image will undergo left-right motion relative to the original image by $n$ times the product of the stride of all convolution layers. Assuming $C=4$ and the sequence shifts four positions to the left, as shown in Fig. \ref{fig3}(c), the positions of the semantic features corresponding to the image features have changed, resulting in a left-right shift in the image content.

Suppose the sequence undergoes a shift of $C \times W \times m$ (where $m$ is a positive integer). In that case, it is equivalent to all semantic features corresponding to different features moving vertically by $m$ positions while maintaining their relative positions horizontally. The decoder can correctly restore the semantic vector. However, the decoded image will undergo vertical motion of $m$ times the product of the stride of all convolutional layers but without horizontal movement. Assuming $C=4$, $W=4$, and the sequence shifts eight positions to the left, as shown in Fig. \ref{fig3}(d), the image content will undergo an up-down shift.

In summary, if the sequence undergoes a shift of $C \times l$ (where $l$ is a positive integer, $l > W$, and $l\% W \neq 0$), the restored image will undergo both left-right and up-down shifts compared to the original image. The pixel distance of the up-down shift is equal to the product of the strides of all convolutional layers in the semantic encoder multiplied by $l//W$, where $l//W$ represents the quotient of $l$ divided by $W$. The pixel distance of the left-right shift is equal to the product of the strides of all convolutional layers in the semantic encoder multiplied by $l\%W$, where $l\%W$ represents the remainder of the division of $l$ by $W$.

\subsection{Semantic Synchronization Method}
This subsection proposes an image-based semantic synchronization method based on the shared SyncImg between transmitter and receiver. This method leverages the inherent semantic information of the SyncImg to realize content-based synchronization localization. At the transmitter, the shared SyncImg is encoded semantically into a one-dimensional vector serving as the synchronization header, followed by the data to be transmitted. As shown in Fig. \ref{fig4}, the receiver employs a sliding window approach for decoding. The window size matches the encoded header length. At each slide of one unit, the data within the window is decoded semantically. The reconstructed SyncImg undergoes classification and template matching for localization:

\begin{figure}[t]
\centering
\includegraphics[width=35mm]{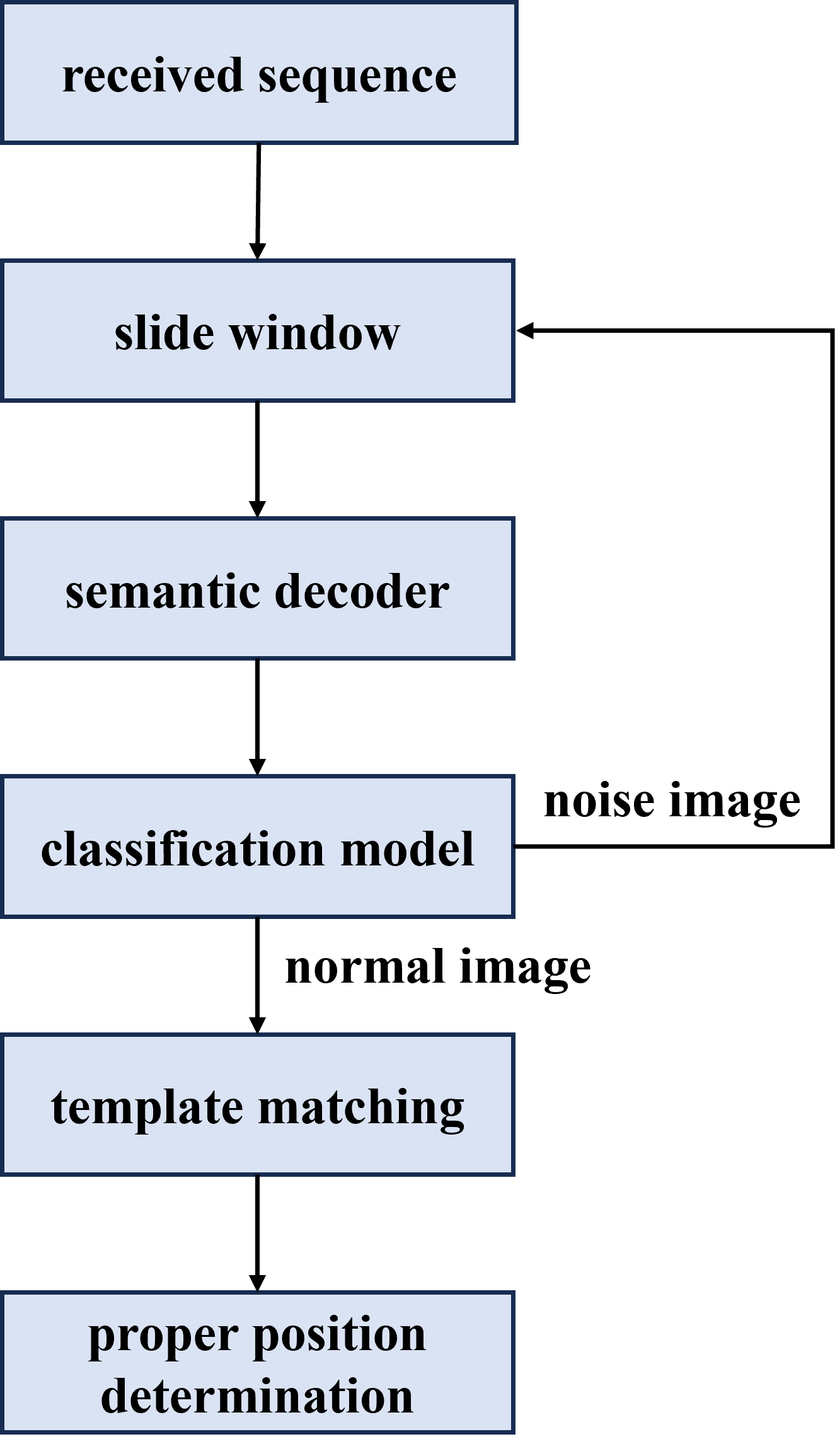}
\caption{Algorithm flow of synchronization module.}
\label{fig4}
\end{figure}

\begin{figure}[ht]
\centering
\includegraphics[width=25mm]{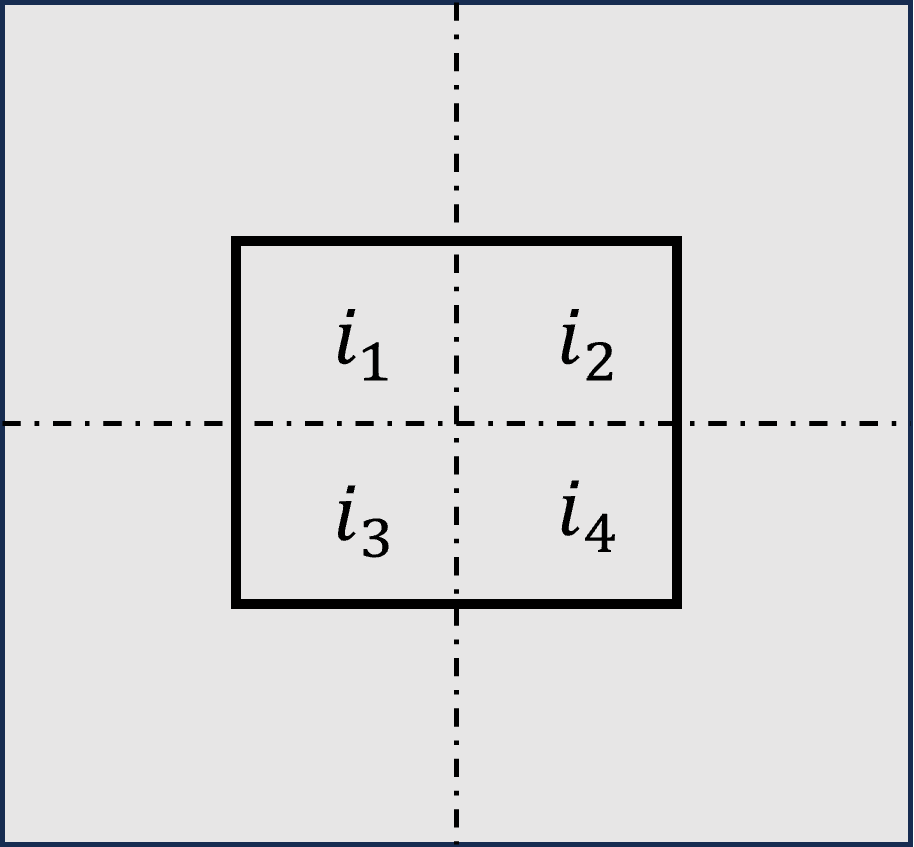}
\caption{Schematic diagram of template matching image block selection.}
\label{fig5}
\end{figure}

\begin{enumerate}
  \item The classification network predicts if the reconstruction is noise, indicating deviation from the proper position by a non-integer multiple of $C$. In this case, the sliding window continues.
  \item If classified as normal, template matching is performed. Four template blocks $i_1$-$i_4$ are extracted centered on the original SyncImg, as shown in Fig. \ref{fig5}. It should be noted that the pixel widths of $i_1$-$i_4$ should be integer multiples of the pixel distance by which the SyncImg content moves when $C$ positions are moved, and the pixel heights are similar. In this way, regardless of how the decoded image shifts, there will always be at least one image block that can be matched with it.
  \item Template matching returns four coordinate pairs $(h'_1,w'_1)-(h'_4,w'_4)$ in the reconstructed image. The offsets from true positions are calculated as follows:
  \begin{equation}
    d_j = \frac{C \times (w'_j - w_j)}{e_1} + \frac{C \times W \times (h'_j - h_j)}{e_2}, 
    \label{eq13}
  \end{equation}
  where $j \in \{1,2,3,4\}$, $e_1$ and $e_2$ are pixel offsets for shifts of $C$ and $C \times W$.
  \item The semantic decoder outputs are obtained at the four offset positions. The SSIM between the decoded image and the original image is calculated. When the SSIM value exceeds the set threshold $\sigma$, the position is determined as the proper synchronization position. 
\end{enumerate}

\section{Experiments}

\subsection{Experiments Setup}
\subsubsection{Semantic Encoder and Decoder Training Dataset}
To reduce the bandwidth occupied by the synchronization header, this paper selects the CIFAR-10 dataset \cite{cifar10} as the training dataset. Considering that the goal of the semantic encoder and decoder is to encode and decode the shared SyncImg between the transmitter and receiver, the generalization ability of the model can be relatively weak. Based on this, this paper randomly selects 1000 images from the CIFAR-10 dataset as the training dataset. Besides, one image among them is specified as the shared SyncImg for both the transmitter and receiver. Due to the training dataset's relative simplicity but the task's specificity, the model will focus more on extracting and reconstructing semantic features of the shared target SyncImg, thus fully meeting the experimental requirements.

\subsubsection{Classification Model Dataset}
The main role of the classification model is to determine whether the current image reconstructed by the decoder is noise or not. Since the transmitter and receiver only share one SyncImg, the generalization ability of the classification model can have a relatively lower requirement. Based on this, for the selected shared SyncImg, this paper performs shifting operations on the semantic encoding sequence of the shared SyncImg to obtain two classes of images for constructing the training dataset of the classification model: images decoded as noise images and images with either left-right or up-down shifted content.

\subsubsection{Structure of the semantic encoder and decoder}
The structure of the semantic encoder and decoder is shown in Table \ref{table1}, where $k$, $s$, $p$, and $f$ represent the kernel size, stride, padding, and number of features, respectively. The semantic encoder and decoder adopt a symmetrical structure design. The main body of the model consists of stacked layers of two-dimensional convolutional networks, where each convolutional layer can extract potential semantic information from the input image at different spatial scales and orientations. Table \ref{table1} provides further details on the parameter settings of each layer's convolutional kernels. The overall structure adopts a down-sampling structure to capture multi-scale semantic features effectively. Two models were trained in this study, one with $c=32$ and the other with $c=16$, corresponding to synchronization header sizes of 1152 symbols and 576 symbols, respectively.
\begin{figure*}[ht]
\centering
\includegraphics[width=\linewidth]{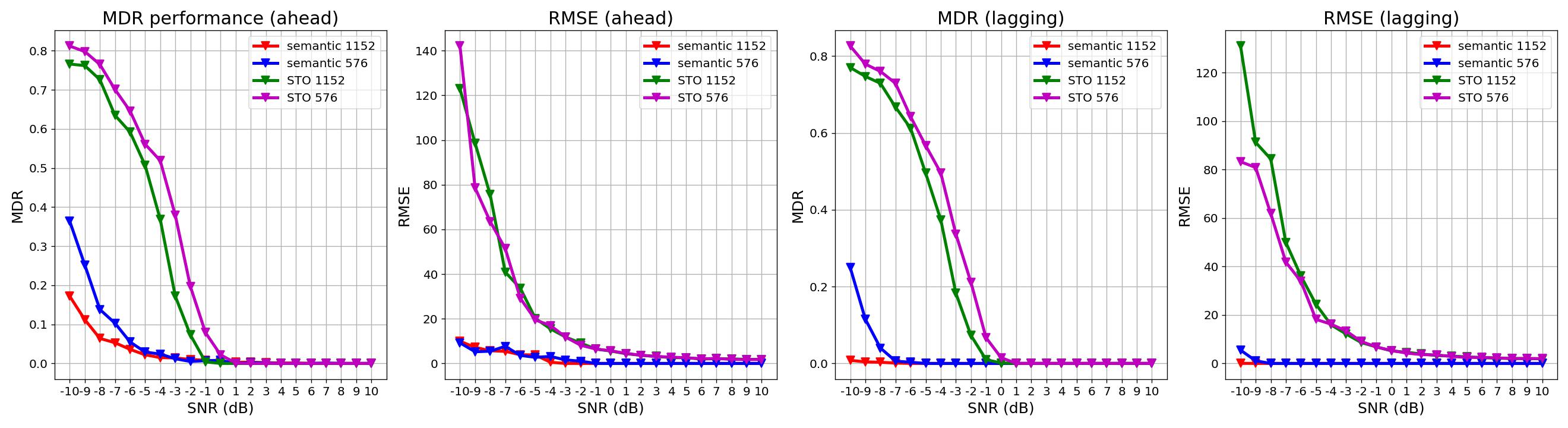}
\caption{The MDR and RMSE performance of the sequences with synchronization header sizes of 1152 and 576 under different SNR conditions when the detection position is three bits ahead of the synchronous position and three bits lagging behind the synchronous position.}
\label{fig6}
\end{figure*}

\begin{figure}[ht]
\centering
\includegraphics[width=\linewidth]{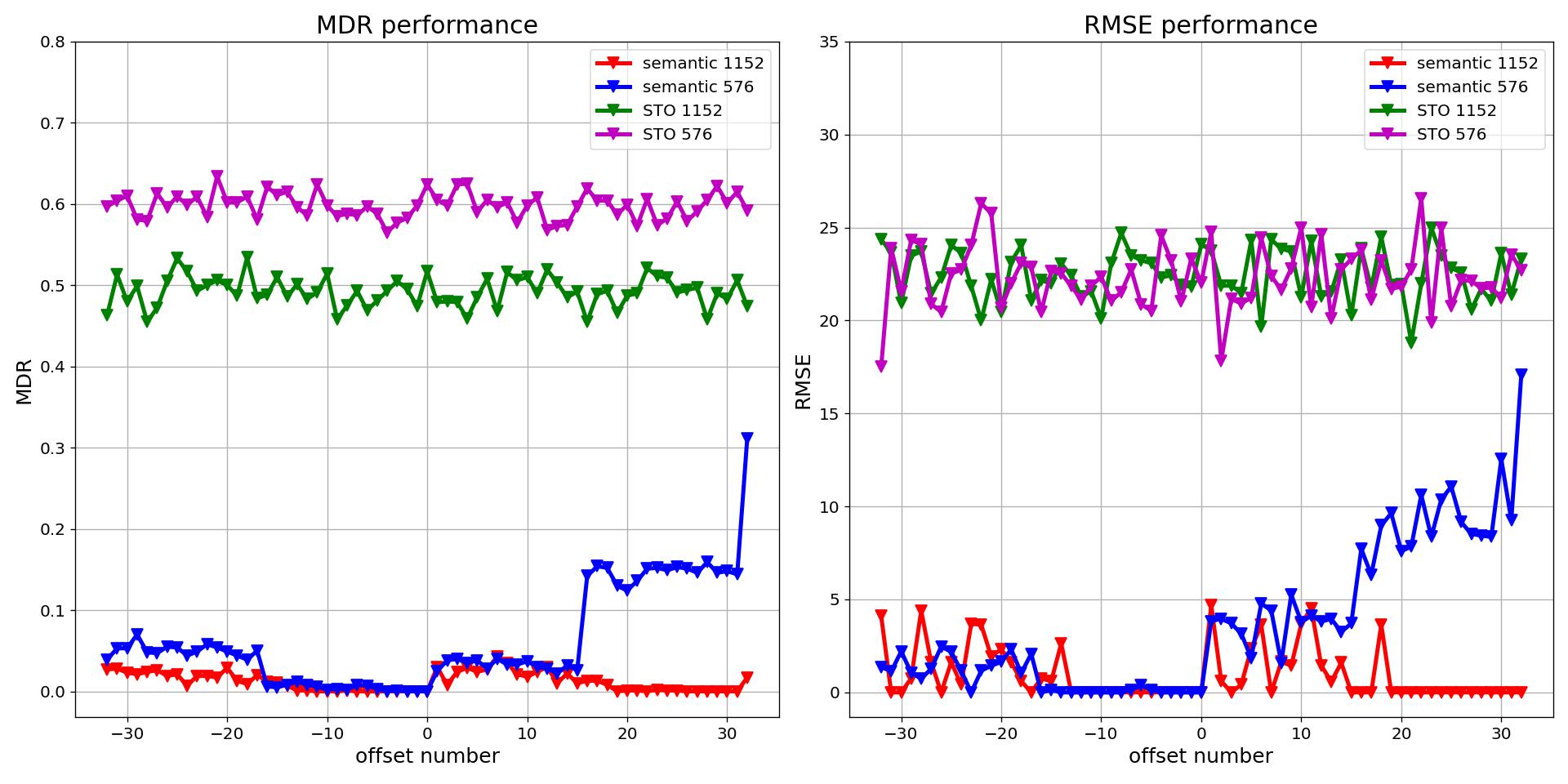}
\caption{The MDR and RMSE performance under different detection positional offsets at an SNR of -5dB.}
\label{fig7}
\end{figure}

\begin{table}[ht]
\caption{Model structure}
\centering
\label{table1}
\begin{tabular}{|c|l|}
\hline
model & \multicolumn{1}{c|}{structure} \\ \hline
\multirow{10}{*}{\begin{tabular}[c]{@{}c@{}} Semantic\\ encoder\end{tabular}} & Conv2d, k=(4,4), s=(2,2), p=(0,0), f=32 \\ \cline{2-2} 
 & Conv2d, k=(3,3), s=(1,1), p=(1,1), f=64 \\ \cline{2-2} 
 & Conv2d, k=(3,3), s=(1,1), p=(1,1), f=128 \\ \cline{2-2} 
 & Conv2d, k=(3,3), s=(1,1), p=(1,1), f=128 \\ \cline{2-2} 
 & Conv2d, k=(3,3), s=(1,1), p=(1,1), f=256 \\ \cline{2-2} 
 & Conv2d, k=(3,3), s=(1,1), p=(1,1), f=256 \\ \cline{2-2} 
 & Conv2d, k=(4,4), s=(2,2), p=(0,0), f=128 \\ \cline{2-2} 
 & Conv2d, k=(3,3), s=(1,1), p=(1,1), f=64 \\ \cline{2-2} 
 & Conv2d, k=(3,3), s=(1,1), p=(1,1), f=32 \\ \cline{2-2} 
 & Conv2d, k=(3,3), s=(1,1), p=(1,1), f=$c$ \\ \hline
\multicolumn{1}{|l|}{\multirow{10}{*}{\begin{tabular}[c]{@{}c@{}}Semantic\\ decoder\end{tabular}}} & Conv2d, k=(3,3), s=(1,1), p=(1,1), f=32 \\ \cline{2-2} 
\multicolumn{1}{|l|}{} & Conv2d, k=(3,3), s=(1,1), p=(1,1), f=64 \\ \cline{2-2} 
\multicolumn{1}{|l|}{} & Conv2d, k=(3,3), s=(1,1), p=(1,1), f=128 \\ \cline{2-2} 
\multicolumn{1}{|l|}{} & ConvT, k=(4,4), s=(2,2), p=(1,1), f=256 \\ \cline{2-2} 
\multicolumn{1}{|l|}{} & Conv2d, k=(3,3), s=(1,1), p=(1,1), f=256 \\ \cline{2-2} 
\multicolumn{1}{|l|}{} & Conv2d, k=(3,3), s=(1,1), p=(1,1), f=128 \\ \cline{2-2} 
\multicolumn{1}{|l|}{} & Conv2d, k=(3,3), s=(1,1), p=(1,1), f=128 \\ \cline{2-2} 
\multicolumn{1}{|l|}{} & Conv2d, k=(3,3), s=(1,1), p=(1,1), f=64 \\ \cline{2-2} 
\multicolumn{1}{|l|}{} & Conv2d, k=(3,3), s=(1,1), p=(1,1), f=32 \\ \cline{2-2} 
\multicolumn{1}{|l|}{} & ConvT, k=(4,4), s=(2,2), p=(1,1), f=3 \\ \hline
\end{tabular}
\end{table}

\begin{table}[ht]
\centering 
\caption{relative execution time}
\label{table2}
\begin{tabular}{|c|c|}
\hline
\multicolumn{1}{|l|}{}  & \multicolumn{1}{l|}{relative execution time} \\ \hline
semantic 576 (ahead)    & 2.63                          \\ \hline
semantic 1152 (ahead)   & 2.46                             \\ \hline
semantic 576 (lagging)  & 1.01                            \\ \hline
semantic 1152 (lagging) & 0.40                            \\ \hline
\end{tabular}
\end{table}

\subsubsection{Structure of the classification model}
The classification model in this paper refers to the Deep Residual Network (ResNet) structure proposed in \cite{resnet}. Specifically, a 101-layer ResNet is adopted as the basic framework.
\subsubsection{Semantic encoder and decoder training details}
An end-to-end learning approach is employed to train the semantic encoder and decoder models. Specifically, the models are trained in an Additive White Gaussian Noise (AWGN) channel environment with an SNR of -3dB. The mean squared error (MSE) between the original image $x$ and the reconstructed image $\hat{x}$ is used as the loss function, as shown below:
\begin{equation}
    \text{loss} = \frac{\sum_{i=1}^{n}(\hat{x}-x)^2}{n}.
    \label{eq10}
\end{equation}
The optimization algorithm uses the Adam optimizer with a learning rate of 0.001. The models are trained for 500 epochs. In the testing stage, the encoding results are first quantized to 3 bits and then transmitted through an AWGN channel with SNR ranging from -10dB to 10 dB.

\subsubsection{Classification model training details}
In this study, a transfer learning approach is employed to train the classification model. Specifically, the weights of the pre-trained ResNet-101 model provided by PyTorch are used for initialization. Then, fine-tuning is performed based on the classification training dataset constructed in this paper. The cross-entropy loss is used as the optimization objective. The Adam optimizer is selected with a learning rate of 0.001. The number of training epochs is 50.

\subsubsection{Baseline}
This paper adopts a time-domain synchronization time offset (STO) estimation technique based on OFDM as the baseline method. The Cyclic Prefix (CP) replicates a portion of the data, i.e., the CP and its corresponding data portion are identical. Considering two sliding windows, $T_1$ and $T_2$, the interval distance between them is the length of an OFDM symbol minus twice the length of the CP. The similarity between the signals within the two windows is calculated, and the maximum likelihood estimation method is used to judge the similarity. When the similarity reaches the maximum value, the current position is determined as the synchronization position. This method sets the carrier frequency offset (CFO) to 0. The CP lengths are set to 1152 and 576, respectively.

\subsubsection{Evaluation metrics}
This paper uses the Miss Detected Ratio (MDR) and Root Mean Square Error (RMSE) as evaluation metrics for synchronization. The MDR is the ratio of the number of frames that fail to detect the correct synchronization position to the total number of transmitted frames, reflecting the success probability of synchronization algorithms. The RMSE is used to measure the difference between the predicted synchronization position $\hat{p}$ and the true synchronization position $p$, and it is calculated as follows:
\begin{equation}
    \text{RMSE} = \sqrt{\frac{\sum(\hat{p} - p)^2}{M}},
    \label{eq11}
\end{equation}
where $M$ represents the total number of transmitted frames. The total number of transmitted frames in the experiments is set to 1000. The threshold of the SSIM is set to 0.6.

\subsection{Result Analysis}
\subsubsection{Synchronization performance}
Fig. \ref{fig6} shows the MDR and RMSE performance of the proposed semantic synchronization method and the traditional OFDM synchronization method under different SNRs when the detection position is three samples ahead of or lagging behind the synchronization position. The results show that in both ahead and behind cases, the semantic synchronization method has lower MDR than the OFDM method at low SNR, and the MDR decreases to zero faster, indicating more robust synchronization locking capability under the same conditions. Furthermore, the proposed semantic synchronization method maintains a low root mean square error (RMSE) performance, implying its favorable synchronization accuracy even under low SNR conditions. Moreover, the effect is better for semantic synchronization when the detection position lags behind the proper position. The reason is that the classification model only needs to run three times to determine the synchronization position, while it needs to run $c-3$ times in the ahead case. Also, the semantic model corresponding to a synchronization header size of 1152 exhibits superior semantic extraction capabilities than the model corresponding to 576, enabling it to exhibit stronger resilience to low SNR conditions.

\par Fig. \ref{fig7} shows the performance of MDR and RMSE under different detection positional offsets from the correct synchronization position at an SNR of -5dB. Negative offset represents detection lagging behind synchronization, while positive offset represents detection leading synchronization. It can be seen that the semantic synchronization method outperforms traditional methods in MDR and RMSE performance. The experiment used the CIFAR-10 image dataset with relatively small image sizes, making template matching challenging when recovery images underwent a left-right shift accurately. For the synchronization header with a length of 576 and channel number 16, when the detection position ahead of synchronization exceeds 16, the classification model detects the recovered SyncImg as normal. Still, the image has undergone a left-right shift, which reduces the matching accuracy and thus degrades synchronization performance. The synchronization header with a length of 1152 and channel number 32 performs better when the detection positional offset is less than 32 from the correct synchronization position.

\subsubsection{Relative execution time}
Table \ref{table2} illustrates the relative execution times of the proposed semantic synchronization method compared to the STO method in OFDM under an SNR of -5dB, when the detection position is three samples ahead of and lagging behind the synchronization position. The relative execution time is defined as the semantic synchronization method's runtime divided by the STO method's runtime. The experiments were conducted using an Intel i9-13900K CPU and NVIDIA A6000 GPU. Due to the larger parameter size of the model corresponding to a synchronization header of 1152 compared to 576, the computational complexity increases, resulting in a longer execution time for the 1152 header model. Additionally, more classification model runs are required when the detection position is ahead of the synchronization position, increasing the computational burden and leading to a relative execution time greater than 1. On the other hand, when the detection position is lagging behind the synchronization position, only three classification model runs are needed, resulting in comparable execution times between the 1152 synchronization header model and the STO method. The synchronization header model with 576 requires less time than the STO method.

\section{Conclusion}
This paper designs a complete semantic synchronization system by leveraging intrinsic information of images and combining it with the semantic communication model. Firstly, this paper theoretically analyzes the impact of semantic signal starting position deviation on image reconstruction. Besides, an image semantic synchronization method is proposed. Experimental results demonstrate that compared with traditional methods, the proposed method achieves lower error rates under low SNR conditions, enabling more precise synchronization tasks. This lays the foundation for reliable transmission of semantic communication in complex channels.

\bibliographystyle{IEEEtran}
\small\bibliography{reference}

\end{document}